# Semantic Temporal Single-photon LiDAR

Fang Li[1*], Tonglin Mu[1*], Shuling Li[1*], Junran Guo[1], Keyuan Li[1],
Jianing Li[1], Ziyang Luo[1], Xiaodong Fan[1], Ye Chen[1], Yunfeng Liu[1],
Hong Cai[2†], Lip Ket Chin[2†], Jinbei Zhang[1†] and Shihai Sun[‡]

[1]*School of Electronics and Communication Engineering, Sun Yat-sen University, Shenzhen, Guangdong, 518107, China*
[2]*Research Institute of Quantum Technologies (RIQT), The Hong Kong Polytechnic University, Hong Kong, China*
*\*These authors contributed equally.*
[†]*corresponding authors: hongcai@polyu.edu.hk; lkchin@polyu.edu.hk; zhjinbei@mail.sysu.edu.cn*
[‡]*chief corresponding author: sunshh8@mail.sysu.edu.cn*

**Abstract:** Temporal single-photon (TSP-) LiDAR presents a promising solution for imaging-free target recognition over long distances with reduced size, cost, and power consumption. However, existing TSP-LiDAR approaches are ineffective in handling open-set scenarios where unknown targets emerge, and they suffer significant performance degradation under low signal-to-noise ratio (SNR) and short acquisition times (fewer photons). Here, inspired by semantic communication, we propose a semantic TSP-LiDAR based on a self-updating semantic knowledge base (SKB), in which the target recognition processing of TSP-LiDAR is formulated as a semantic communication. The results, both simulation and experiment, demonstrate that our approach surpasses conventional methods, particularly under challenging conditions of low SNR and limited acquisition time. More importantly, our self-updating SKB mechanism can dynamically update the semantic features of newly encountered targets in the SKB, enabling continuous adaptation without the need for extensive retraining of the neural network. In fact, a recognition accuracy of 89% is achieved on nine types of unknown targets in real-world experiments, compared to 66% without the updating mechanism. These findings highlight the potential of our framework for adaptive and robust target recognition in complex and dynamic environments.

## 1. Introduction

Targeting imaging and recognition in complex environments and adverse conditions (such as rain, fog, long distances, weak echo signals, *etc.*) is a fundamental yet challenging task for many applications, including autonomous vehicle navigation [1,2], air defense [3] and remote sensing [4,5]. Single-photon LiDAR, which sends a pulsed laser and detects the echo with a single-photon detector (SPD), offers potential advantages for precise imaging and recognition in these applications, as it possesses picosecond-level temporal resolution and single-photon sensitivity. Many single-photon LiDAR systems, including SPD arrays imaging [6], scanning SPD imaging [7–9], ghost imaging [10–12] and computational SPD imaging [13], have been demonstrated. These systems typically focus on generating high-quality images, followed by the extraction of extra target features and classification algorithms with the help of machine learning [14–16]. Although these systems perform well in many cases, they face challenges in terms of speed, power consumption, and system complexity.

Recently, a temporal single-photon LiDAR (TSP-LiDAR) was proposed and demonstrated [17–21], which focuses on directly extracting target features from the 1D time-of-flight temporal histogram of the echo single-photon signal from the target to the SPD by using an AI algorithm and a pre-trained neural network [22–24]. Then, the precious imaging phase can be bypassed, and high performance with high speed and a simple system structure can be achieved [25–27]. However, current TSP-LiDAR still faces two main bottlenecks. First, weak conditions, such as low signal-to-noise ratio (SNR) and short acquisition time (resulting in fewer echo photons), will compromise its accuracy and speed. Second, it will be invalid for open-set scenarios. The targets that need to be reconstructed or identified should be known in the pre-trained network. If an unknown target appears in practical cases, it will be misclassified and the neural network should be retrained accordingly.

In this paper, to overcome the two bottlenecks, a TSP-LiDAR based on a self-updating semantic communication model, referred to as the semantic TSP-LiDAR, is proposed and demonstrated. Instead of faithfully transmitting physical symbols (the raw message) between two parties, semantic communication focuses on conveying the semantic features of the message. The receiver then reconstructs the original message by matching the received semantic features with a pre-shared semantic knowledge base (SKB) containing the features of all possible messages. This approach can significantly enhance channel capacity [28–32]. Here, we show that the TSP-LiDAR can be modeled as the semantic communication, since they have the same signal processing pipeline. The detection phase involves illuminating the target with a pulsed laser and detecting a single-photon echo signal from the target, which can be considered a semantic feature encoding. In this process, the 3D geometric shape of the target (raw message) is mapped into a 1D temporal histogram (semantic feature). The target reconstruction or identification phase is equivalent to the semantic matching processing, in which the target information is extracted from the detected 1D temporal histogram.

The results, from both the indoor experiment and the simulation, show that our semantic TSP-LiDAR performs well under shorter acquisition times and worse SNR conditions. For instance, with a short acquisition time of 10 ms, our semantic TSP-LiDAR achieves an average accuracy exceeding 55%, much higher than the current TSP-LiDAR, which is less than 30% for ResNet-18 and 20% for U-Net. Furthermore, by designing an SKB self-updating mechanism, our semantic TSP-LiDAR could also identify unknown targets very well. For the experimental dataset, as the number of unknown classes increases to 9, the recognition accuracy of the baseline without the self-updating mechanism declines to 66%. In contrast, the proposed self-updating mechanism maintains an accuracy of 89%. For the synthetic dataset, when the number of unknown classes reaches 10, the accuracy without the mechanism decreases to 79%, whereas the self-updating mechanism maintains an accuracy of 95%. These results underscore the significant potential of our semantic TSP-LiDAR for accurate and adaptive target recognition in challenging real-world scenarios, offering a substantial advancement over existing single-photon LiDAR systems.

## 2. Method

### 2.1 Semantic model of TSP-LiDAR

Figure 1(a) illustrates the principle of TSP-LiDAR. A pulsed laser illuminates the scene, and the echo photons are detected by an SPD and recorded by a time-to-digital converter (TDC). The arrival times of photons are accumulated into a temporal histogram, which is then processed by a pre-trained neural network to identify the target based on the extracted features. Figure 1(b) illustrates the equal semantic model of TSP-LiDAR, where we reformulate the process of TSP-LiDAR target recognition as a semantic communication. By constructing a shared SKB and transmitting only the task-relevant information, semantic communication could improve both transmission efficiency and noise robustness. Thus, our semantic TSP-LiDAR also performs well with unknown objects and under worse conditions, which consists of three main stages: encoding, transmission, and decoding.

The object illumination process serves as the semantic encoding stage, where the 3D spatial information of the object is projected into a 1D temporal domain. A pulsed laser $p(t)$ flood-illuminates the object $o$, and the geometric shape and reflectivity index of the object determine the temporal distribution of the echo photons. This processing can be considered as a map from the spatial to the temporal domain,

$$s(t) = f(o), \tag{1}$$

where $s(t)$ denotes the ideal temporal histogram and $f(\cdot)$ represents the physical transformation governed by the target's structural and optical properties.

During the transmission stage, the transmitted signal $s(t)$ undergoes distortion through the channel due to imperfections of the detection system. Specifically, two primary noise sources affect the received signal $x(t)$. One is the jitter of detectors, which refers to timing uncertainty in the detection system. Another one is background noise, including the background light and the dark count of the SPD. These noise sources together degrade the fidelity of the received histogram, especially in photon-limited conditions.

The detection and recognition process corresponds to the decoding stage. In traditional schemes, the receiver attempts to make predictions of the class by finding an optimal inverse mapping $f^{-1}$,

$$y_T = f^{-1}[x(t)]. \tag{2}$$

In contrast, the semantic TSP-LiDAR framework focuses on extracting object-relevant semantic feature $k_R$ from the received histogram, which can be considered a mapping $f_S$ from $x$ to $k_R$. It is not that, since the complexity of $f_S$, the extraction of object features is implemented with a pre-trained neural network. Then the extracted semantic feature is compared with the entries in the SKB to find the best match and then classify the object. The predicted class label in the semantic matching process can be expressed as

$$y_S = \underset{m \in \mathcal{M}}{\mathrm{argmin}}\, \mathcal{D}(k_R = f_S[x], k_m), \tag{3}$$

where $k_m$ denotes the semantic feature of the $m$-th class in the SKB; $\mathcal{M} = \{1,2,\ldots,M\}$ denotes the set of classes in the SKB, and $\mathcal{D}(\cdot)$ is a match function used to calculate the distance between $k_R$ and $k_m$, which is defined as

$$\mathcal{D}(k_R, k_m) = \frac{k_R \cdot k_m}{\parallel k_R \parallel \parallel k_m \parallel}, \tag{4}$$

where $\parallel \cdot \parallel$ represents the Euclidean norm.

By interpreting TSP-LiDAR as an equal semantic model, the recognition task becomes a semantic-matching problem. This reformulation enables the system to utilize the SKB for efficient, noise-resilient, and adaptive classification or image reconstruction. Moreover, it lays the foundation for self-updating and open-set recognition in subsequent stages, significantly improving the system's scalability and robustness under limited acquisition time and low SNR conditions.

## 2.2 Self-updating mechanism

Another important advantage of our semantic TSP-LiDAR is that a self-updating mechanism could be introduced to enable adaptive target recognition in open-world environments. This mechanism enables the SKB to evolve dynamically during inference, allowing for the recognition of both known and previously unknown objects without requiring retraining of the neural network. The overall framework of the semantic TSP-LiDAR with a self-updating mechanism is illustrated in Fig. 2(a), which follows semantic feature extraction, unknown detection, and SKB update or semantic matching.

The semantic feature extraction plays a crucial role in our semantic TSP-LiDAR, which is achieved through a probabilistic encoder–decoder network [33] called STSP Net, as shown in Fig. 2(b). In this network, the encoder outputs the mean and variance parameters $(\mu, \sigma^2)$, from which a latent vector is then sampled using the reparameterization trick [34]. The decoder then reconstructs the input histograms from the sampled vector, ensuring that the learned features preserve meaningful and informative structures. To enhance intra-class semantic consistency, each class $m \in \mathcal{M}$ is assigned a class-specific semantic center $\mu_m$, derived from the one-hot encoded label through a fully connected mapping. The overall training loss contains two terms: the reconstruction loss $L_{rec}$, which ensures accurate histogram reconstruction, and the Kullback-Leibler divergence loss $L_{KL}$ [35], which enforces class-level semantic separability. After training, the mean vector $\mu$ output by the encoder of STSP Net serves as a semantic feature. Then, the semantic features belonging to the same class $m$ are averaged to obtain the class-level semantic feature $k_m$, which is then used to construct the initial SKB $K_0$. The detailed procedures for model training and SKB construction are provided in the supplementary material. During inference, the semantic feature $z = (z_1, \cdots, z_d)$ of the received temporal histogram $x(t)$ is extracted directly from the encoder of the trained STSP Net.

To detect potential unknown targets, a likelihood-based detection strategy is employed. Each known class $m \in \mathcal{M}$ in the SKB is modeled as a multivariate Gaussian distribution $f_m(z) = N(z; k_m, \Sigma_m)$, where $k_m$ is the semantic feature of $m$-th class and $\Sigma_m = \text{diag}(\sigma_m^2)$ is the diagonal covariance matrix. Both parameters are estimated in the SKB construction stage. Then, the likelihood of a semantic feature $z$ locating in the distribution of the $m$-th class is expressed as

$$P_m(z) = 1 - \prod_{i=1}^{d} \int_{k_{m,i}-|z_i-k_{m,i}|}^{k_{m,i}+|z_i-k_{m,i}|} f_m(t_i) \mathrm{d}t_i, \tag{5}$$

where $k_{m,i}$ is the $i$-th dimension of $k_m$. Intuitively, the farther $z$ is from $k_m$, the smaller the value of $P_m(z)$, indicating a lower likelihood that $z$ belongs to the class $m$.

When none of $P_m(z)$ surpass the threshold $\tau$[30], the sample is categorized as an unknown class with a semantic feature defined as

$$k_{\text{new}} = \frac{1}{N} \sum_{j=1}^{N} z^{(j)}, \tag{6}$$

where $N$ is the number of samples identified as the same unknown class and $z^{(j)}$ is the extracted semantic feature of the $j$-th sample. The new semantic feature $k_{new}$ is then added to the SKB, updating it from its initial state $K_0$ to $K_1 = K_0 \cup \{k_{M+1} = k_{new}\}$. This self-updating mechanism enables the system to incorporate previously unknown categories without requiring model retraining.

Conversely, when $P_m(z)$ exceeds a threshold $\tau$ for any $m \in \mathcal{M}$, the sample is identified as a known target, and semantic matching is applied for classification. The class with the highest similarity (see Eq. 3) is selected as the classification result, and the predicted label $y$ is calculated as

$$y = \underset{m \in \mathcal{M}}{\text{argmin}}\, \mathcal{D}(z, k_m). \tag{7}$$

Through the integration of semantic feature extraction and dynamic SKB evolution, the proposed self-updating mechanism allows the semantic TSP-LiDAR system to achieve robust, noise-resilient, and continuously evolving target recognition.

## 3. Results and Discussion

### 3.1 Experimental system and dataset

To evaluate the performance of the proposed semantic TSP-LiDAR and self-updating recognition framework, comprehensive experiments were conducted. The experimental dataset was collected using a homemade TSP-LiDAR system, as shown in Fig. 3(a). A pulsed laser with a central wavelength of 1550 nm, an average power of approximately 200 mW, a time width of 10 ps, and a repetition rate of 20 MHz is generated to illuminate the object, which is placed at a distance between 2 and 3 m from the laser. The echo photons from the target are detected by a free-running SPD (QCD600B from QuantumCTek), which features a dead time of 900 ns and a photon detection efficiency of 30%. A TDC with a bin width of 10 ps is adopted to record the clock of APD, which is synchronized with the laser, ensuring precise photon arrival time measurements. Then all data are updated to a laptop for further data processing.

Different acquisition times (1, 2, 5, 7, 10, 50, 100, 500, and 1000 ms) are set to obtain various counts of echo photons, enabling the evaluation of the system's performance under varying photon flux conditions. The detected signal photon number ranges from single digits to several thousand, when the acquisition time increases from 1 ms to 1 s. In the experiment, we test 39 classes in total, including 9 distinct objects with four different angles (front, back, left, and right) and three background classes. For each class, 200 samples are captured. Fig. 3(b) presents the front view of the 9 target objects (other categories and corresponding histograms are provided in the supplementary material).

To assess the baseline classification performance, the unknown categories are first ignored and all 39 classes are included in the training set. We compare the performance of our semantic TSP-LiDAR with that of two commonly used deep learning models, namely ResNet-18 [23] and U-Net [22], under varying acquisition times. The sampled data in experiments is partitioned into training, validation, and testing sets with a ratio of 7:1:2. For both ResNet-18 and U-Net, the softmax cross-entropy loss function is employed to train both networks. The received signal is then directly fed into the networks, which output the predicted class labels. For our semantic TSP-LiDAR, the validation set is utilized to construct the SKB, which serves as the foundation for semantic matching during the testing phase. We then conduct open-set tests, in which only 30 categories are included in the training set and are treated as known. We first hold out one category as unknown and compare performance with and without the self-updating mechanism across different acquisition times. We then fix the acquisition time at 1000 ms and gradually increase the number of unknown categories up to 9 to further investigate the effect of the number of unknown classes.

### 3.2 Synthetic dataset

To further verify the performance of our semantic TSP-LiDAR, a large-scale synthetic dataset was generated following the physically grounded simulation model [17]. This dataset enables controlled evaluation under a wide range of SNR conditions.

The dataset comprises 50 object categories, each derived from 3D models in publicly available repositories such as ShapeNet. For each model, temporal histograms are simulated by projecting the 3D object onto 2D depth and reflective maps (pixel values are set to 1 for target pixels and zero otherwise), followed by photon-level modeling of the ToF process. Both dark counts and ambient illumination were incorporated as Poisson-distributed noise sources, capturing the stochastic nature of single-photon detection. Simulations were conducted under 26 SNR levels, ranging from $-19.19$ dB to $13.01$ dB, with the total photon count per sample fixed at $2 \times 10^7$. For each SNR level, 1000 samples per category were generated, resulting in a statistically robust dataset for evaluating recognition accuracy under varying noise conditions.

Similar to the experimental dataset, we first exclude unknown categories and include all 50 classes in the training set to evaluate the baseline classification performance on the synthetic dataset. Subsequent data split and training procedure are the same as those used for the experimental dataset. Then, the open-set performance with 40 categories treated as known is assessed. One category is initially treated as unknown, and performance is compared both with and without the self-updating mechanism across varying SNR conditions. Next, the SNR is fixed at 13.01 dB and the number of unknown categories is progressively increased up to 10 to examine sensitivity to the size of the unknown set.

*3.3 Results*

Figure 4(a) shows the classification accuracy versus acquisition time. It is evident that our semantic TSP-LiDAR outperforms both ResNet-18 and U-Net. At shorter accumulation times (e.g., below 10 ms), our semantic TSP-LiDAR exhibits a significant advantage, achieving higher accuracy. As the acquisition time increases (e.g., beyond 50 ms), the accuracy of all methods converges to a similar level, while the proposed method still maintains a slight edge. The classification performance on the synthetic dataset is shown in Fig. 4(b). The proposed method demonstrates superior robustness to noise, achieving higher accuracy than ResNet-18 across a wide range of SNR values. At an SNR of $-16$ dB, our semantic TSP-LiDAR outperforms ResNet-18 and U-Net by 28% and 48%, respectively, in terms of classification accuracy.

Figure 5(a) shows the performance with and without SKB update under varying acquisition times and SNRs. With the help of self-updating, an improvement in accuracy is observed, confirming the model's ability to adapt dynamically to unknown targets. The result when increasing the number of unknown classes is shown in Fig. 5(b). For the experimental dataset, the accuracy of the non-updated SKB decreases sharply to about 66% when the number of unknown categories increases to 9. In contrast, the updated SKB maintains an accuracy of 89%. For the synthetic dataset, when the number of unknown categories increased to 10, the accuracy of the non-updated model declined to 79%, while the updated SKB maintained an accuracy of 95%. With the SKB update, although minor fluctuations were observed as the number of unknown classes increased, the overall accuracy remained stably above 89%. These results confirm that the proposed semantic TSP-LiDAR with a self-updating SKB achieves robust open-set recognition, effectively incorporating unknown targets without retraining, and remaining resilient under severe photon noise.

## 4. Conclusions

This paper demonstrates a self-updating SKB-enabled semantic framework for TSP-LiDAR target recognition. By reformulating the LiDAR sensing and inference process as an equal semantic model, the proposed system focuses on task-relevant semantic representations, thereby improving robustness against photon scarcity and noise. A self-updating mechanism is further introduced, enabling the SKB to evolve during inference through likelihood-based unknown detection and adaptive knowledge integration, without retraining the model.

Comprehensive evaluations on both experimental and synthetic datasets verify the superiority of the proposed approach. Even under challenging conditions, such as short acquisition time (10 ms) and severe noise level (SNR $= -16$dB), the method maintains classification accuracy above 55%. Moreover, the self-updating mechanism effectively preserves recognition performance when unknown categories appear. Without the self-updating mechanism, accuracy decreases to 66% for experimental targets and 79% for synthetic targets as the number of unknown classes increases to 9 and 10, respectively. In contrast, the proposed approach consistently maintains an average accuracy above 89% in both cases. Therefore, this self-updating SKB-enabled semantic approach provides a robust and scalable solution for TSP-LiDAR target recognition in dynamic and open-set environments. Its ability to continuously refine the SKB and improve recognition accuracy under various conditions makes it well-suited for real-world applications, such as autonomous systems and remote sensing. In particular, the framework is compatible with resource-constrained and edge-deployed platforms, where retraining is costly or infeasible, and reliable perception must be maintained under photon-limited and noisy conditions. The self-updating mechanism also provides a principled way to incorporate emerging object categories, which is critical for long-term operation in safety-critical domains such as intelligent transportation, environmental monitoring, and infrastructure inspection. Beyond TSP-LiDAR, the proposed semantic modeling and SKB update strategy can be extended to other time-resolved or photon-limited sensing modalities, offering a general paradigm for building adaptive, knowledge-driven perception systems.

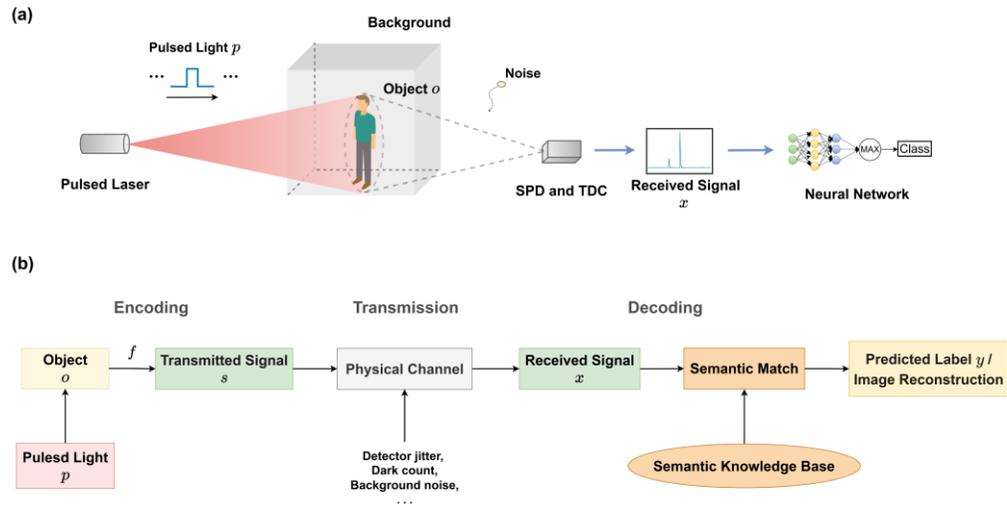

Fig. 1. (a) Schematic of the traditional target recognition process using TSP-LiDAR, where the temporal histogram of the received single-photon signal is directly analyzed by a neural network for classification. (b) The proposed equal semantic model of TSP-LiDAR establishes an equivalence between the TSP-LiDAR signal processing pipeline and a semantic communication system. The encoding, transmission, and decoding stages correspond, respectively, to laser illumination and reflection, photon detection under noise, and semantic matching based on a shared SKB for target recognition or image reconstruction.

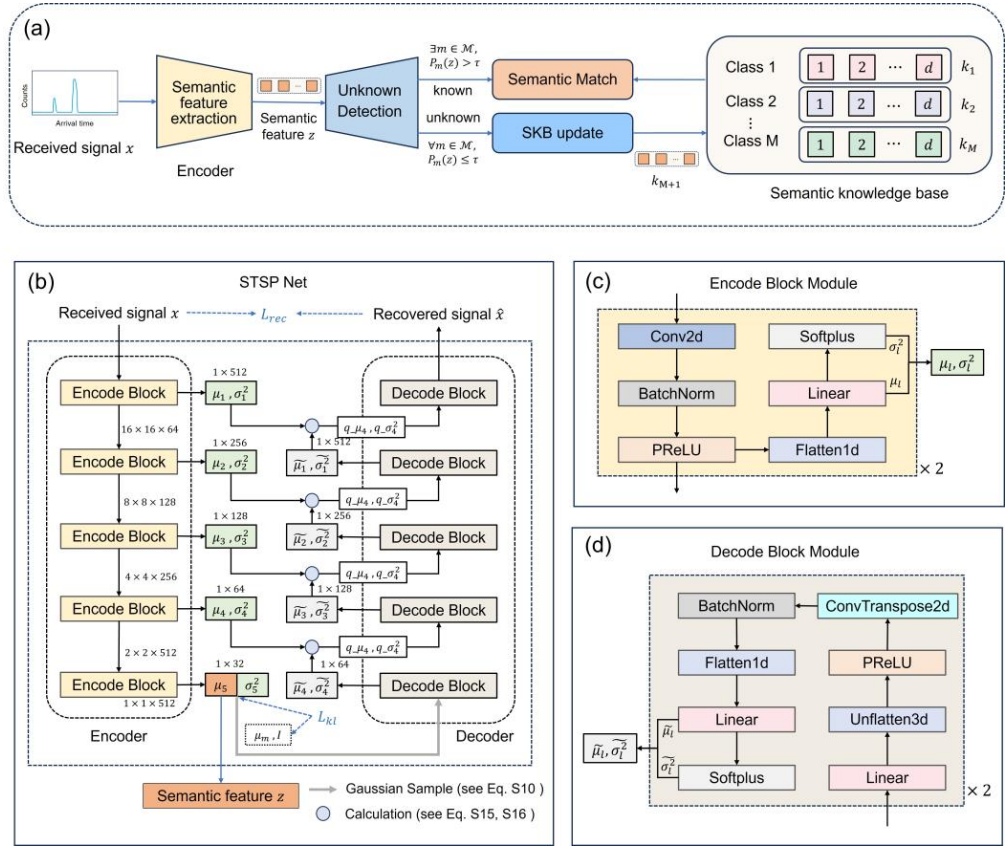

Fig. 2. (a) Workflow of the semantic TSP-LiDAR system with a self-updating mechanism. The model extracts semantic features from temporal histograms, detects unknown samples, and updates the SKB to enable open-set, self-adaptive recognition. (b) Internal architecture of the STSP Net. (c) and (d) show the detailed structure of the encode block and decode block of the STSP Net, respectively.

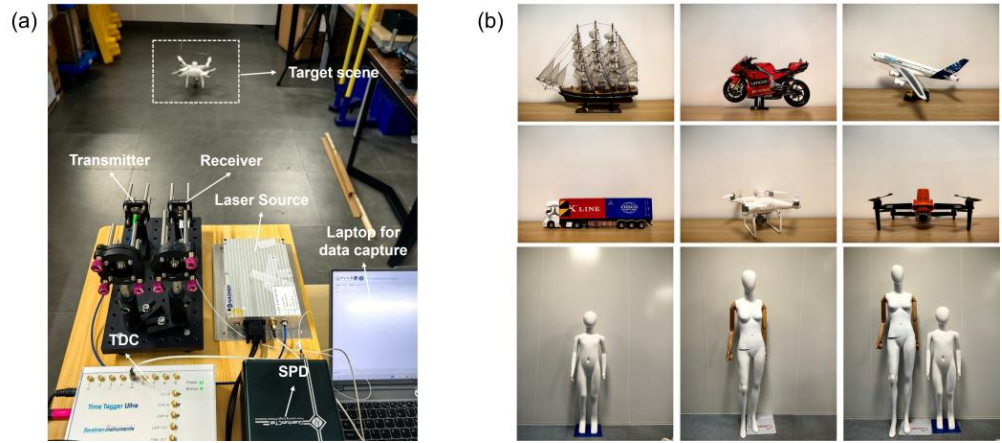

Fig.3 (a)TSP-LiDAR system for experimental dataset capture. The target shown in (a) is the class "DJI UAV". (b) 9 classes of the experimental dataset, which are the front of a boat, motorbike, plane, truck, DJI UAV, UAV, child, adult, and two people.

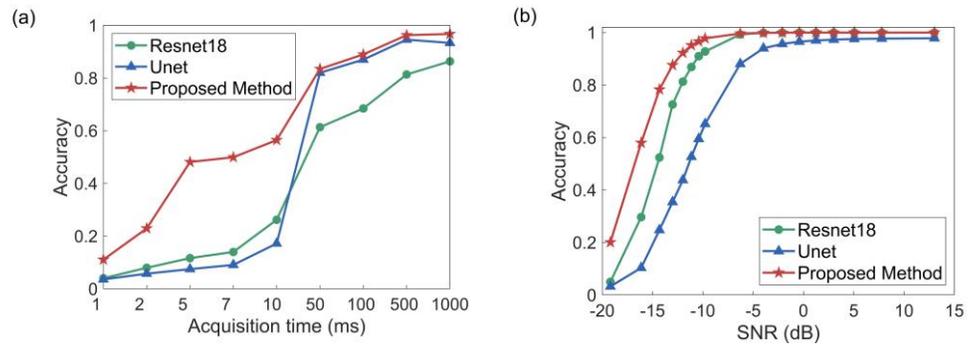

Fig.4 Comparison of different methods, including ResNet-18, U-Net, and the proposed semantic TSP-LiDAR-based SKB. The unknown categories are excluded from the test set. Results are shown for (a) the experimental dataset and (b) the synthetic dataset.

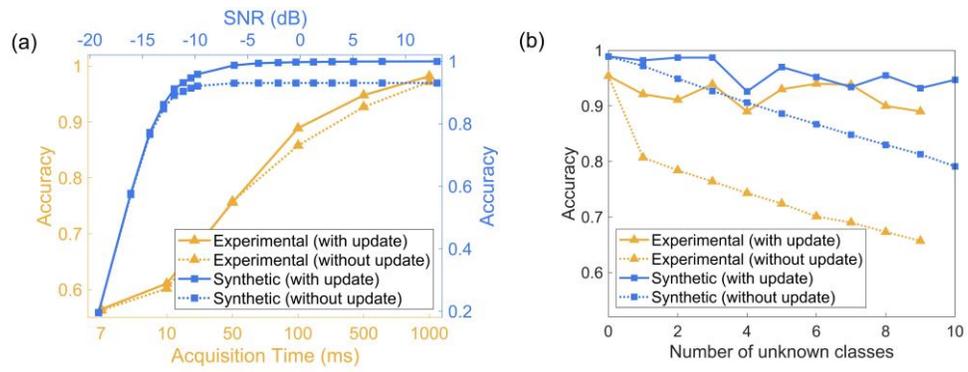

Fig.5 Comparison of SKB update effects on experimental (orange) and synthetic (blue) datasets with unknown classes. Solid lines represent results with the SKB update, while dotted lines represent results without the update. (a) Accuracy versus acquisition time and SNR with one unknown class. (b) Accuracy versus the number of unknown classes.